\documentclass[prd,amsmath,amssymb,superscriptaddress,preprintnumbers,twocolumn,nofootinbib,10pt]{revtex4-1}
\usepackage{amsmath,amssymb,graphics,epsfig,subfigure}
\usepackage{color}
\usepackage{mathrsfs}

\pdfoutput=1
\usepackage{graphicx}
\usepackage{dcolumn}
\usepackage{bm}
\usepackage{amssymb}
\usepackage{latexsym}
\usepackage{booktabs}
\usepackage{amsmath}
\usepackage{multirow}
\usepackage{url}
\usepackage{footnote}
\usepackage{float}
\usepackage{threeparttable}
\usepackage[colorlinks=true, linkcolor=blue, citecolor=blue]{hyperref}
\usepackage[bottom]{footmisc}

\usepackage[normalem]{ulem}
\usepackage{color}
\usepackage{array}
\usepackage{enumerate}
\usepackage{adjustbox}

\usepackage{makecell}
\usepackage{diagbox}
\usepackage{epstopdf}
\usepackage{epsfig}
\usepackage{longtable}
\usepackage{supertabular}
\usepackage{algorithm}
\usepackage{pifont}
\usepackage{algorithmic}
\usepackage{changepage}
\usepackage{setspace}

\usepackage{CJKutf8}

\begin{document}
\begin{CJK*}{UTF8}{gbsn}

\title{Characterized behaviors of black hole thermodynamics in the supercritical region}

\author{Zi-Qiang Zhao(赵子强)}
\affiliation{Key Laboratory of Cosmology and Astrophysics (Liaoning), College of Sciences, Northeastern University, Shenyang 110819, China}

\author{Zhang-Yu Nie(聂章宇)}
\email{niezy@kust.edu.cn}
\affiliation{Center for Gravitation and Astrophysics, Kunming University of Science and Technology, Kunming 650500, China}



\author{Jing-Fei Zhang(张敬飞)}
\affiliation{Key Laboratory of Cosmology and Astrophysics (Liaoning), College of Sciences, Northeastern University, Shenyang 110819, China}

\author{Xin Zhang(张鑫)}
\email{zhangxin@neu.edu.cn}
\affiliation{Key Laboratory of Cosmology and Astrophysics (Liaoning), College of Sciences, Northeastern University, Shenyang 110819, China}
\affiliation{Key Laboratory of Data Analytics and Optimization for Smart Industry (Ministry of Education), Northeastern University, Shenyang 110819, China}
\affiliation{National Frontiers Science Center for Industrial Intelligence and Systems Optimization, Northeastern University, Shenyang 110819, China}

\begin{abstract}
\noindent\textbf{Abstract:} The comprehension of universal thermodynamic behaviors in the supercritical region is crucial for examining the characteristics of black hole systems under high temperature and pressure. This study is devoted to the analysis of characteristic lines and crossover behaviors within the supercritical region. By making use of the free energy, we introduce three key thermodynamic quantities: scaled variance, skewness, and kurtosis. Our results demonstrate that the Widom line, associated with the maximal scaled variance, can effectively differentiate between small and large black hole-like subphases, each displaying distinct thermodynamic behaviors within the supercritical region. Furthermore, by utilizing quasinormal modes, we identify the Frenkel line, offering a dynamic perspective to distinguish between small and large black hole-like subphases. These contribute to a deeper comprehension of black hole subphases in the supercritical region, thus illuminating new facets of black hole thermodynamics.


\end{abstract}


\maketitle

{\emph{1. Introduction}.} Black hole thermodynamics has been a pivotal area of study since the groundbreaking contributions of Bekenstein and Hawking~\cite{Bekenstein73,Hawking75}. The surface gravity and area of the event horizon are related to the black hole's temperature and entropy, offering insights into these mysterious objects of extreme gravity. This also prompts us to reconsider the nature of black holes.

A phase transition is a remarkable occurrence in conventional thermodynamic systems, akin to the transitions observed among the states of water, liquid, ice, and gas that we encounter in our daily lives. Remarkably, a similar phenomenon manifests within black holes. However, unlike typical systems, the free energy of a black hole system can be determined by assessing the Euclidean action through a gravitational path integral in a ``zero-loop" approximation \cite{Gibbons77}
\begin{eqnarray}
 \mathcal{Z}=e^{-\beta \mathcal{F}}=\int D[g]e^{-\frac{\mathcal{I}}{\hbar}},\label{action}
\end{eqnarray}
where $\mathcal{Z}$, $\mathcal{F}$, and $\mathcal{I}$ represent the partition function, free energy, and Euclidean action of the black hole with a temperature $T=1/\beta$. Through the analysis of free energy, Hawking and Page initially demonstrated the presence of a phase transition between the pure thermal radiation phase and the stable large black hole phase in AdS space~\cite{HawkingPage1983}. This discovery, motivated by the AdS/CFT correspondence, effectively interprets the confinement/deconfinement phase transition of gauge fields~\cite{Witten1998-2}. Such discovery has sparked keen interest in exploring phase transitions in gravitational systems and in analogy to quantum chromodynamics (QCD) systems.

Moreover, it has been demonstrated that for charged AdS black holes, the free energy exhibits swallowtail behaviors, signifying the presence of phase transitions between small and large black hole phases~\cite{Chamblin}, which are reminiscent of van der Waals (VdW) fluids. This observation was further confirmed in Ref.~\cite{Kubiznak2012}, where critical phenomena were identified by interpreting the cosmological constant as pressure in the extended phase space~\cite{Kastor2009}. Through the analysis of free energy, various intriguing types of phase transitions have been unveiled.

In addition to studying phase transitions, the free energy has been utilized to develop Ruppeiner geometry~\cite{Ruppeiner},  with the aim of understanding the intrinsic microstructure of black holes.~\cite{Wei20152016,Wei2019,Wei2024}. Furthermore, the establishment of thermodynamic topology has been achieved through the use of generalized free energy~\cite{WeiDefect}. Hence, free energy plays a pivotal role in investigating the thermodynamics and phase transitions of black holes.

Previous studies have predominantly concentrated on investigating the thermodynamic properties of phase transitions below and near the critical point, with challenges arising when differentiating between the small and large black hole phases beyond this threshold. However, recent studies have unveiled that within this supercritical region, gas-like and liquid-like subphases emerge not only in classical systems like water~\cite{Brazhkin2013,Gallo2014} but also in microscopic systems such as QCD~\cite{Stephanov2011,Vovchenko2015}, shedding light on the underlying thermodynamic phenomena. We will provide criteria for different types of supercritical black holes from both thermodynamic (Widom line) and dynamic (Frenkel line) perspectives. In thermodynamics, we will primarily use free energy as the criterion, while in dynamics, we will mainly use quasinormal modes (QNMs) as the criterion.


As demonstrated earlier, free energy serves as a crucial thermodynamic parameter for exploring phase transitions below and near the critical points. It is anticipated to play a significant role in the supercritical region as well. Inspired by the research on QCD~\cite{Vovchenko2015}, we introduce the scaled variance $\Omega$, skewness $S\sigma$, and kurtosis $\kappa\sigma^2$ as part of our study,
\begin{eqnarray}
\Omega=\frac{k_2}{k_1},\quad  S\sigma=\frac{k_3}{k_2},\quad \kappa\sigma^2=\frac{k_4}{k_2}, \label{qqq}
\end{eqnarray}
making use of the derivatives of the Gibbs free energy
\begin{eqnarray}
k_n \equiv \frac{\partial^n G(T,P)}{\partial T ^n}.
\end{eqnarray}
We shall show that these quantities (\ref{qqq}) can indeed reflect the properties of the black hole in supercritical region as expected.

\begin{figure}
	\label{phaseDiagram}\includegraphics[width=6cm]{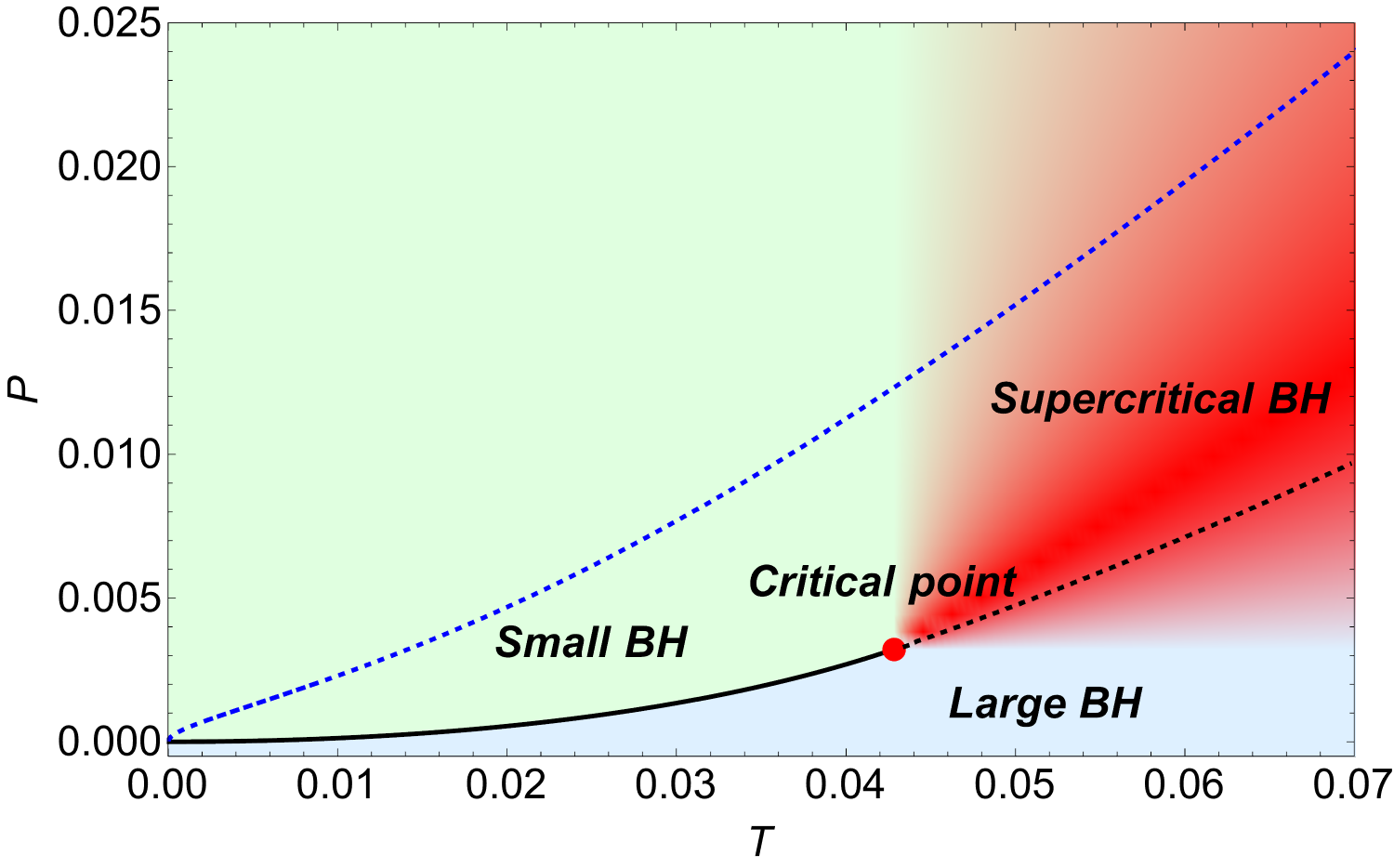}
	\caption{Phase diagram for the charged AdS black hole. The black solid curve is the coexistence curve of the small and large black holes. The black dashed line represents the widom line, and the blue dashed line represents the Frenkel line. These lines divide the supercritical region into the small and large black hole-like subphases from the perspective of thermodynamics and dynamics, respectively.}\label{pphaseDiagram}
\end{figure}

{\emph{2. Widom line and crossover behavior}.} The Widom line stands out as a distinguished boundary demarcating the transition between gas-like and liquid-like (or small and large black hole-like) subphases within the supercritical region. To exemplify this, we consider the four-dimensional charged AdS black hole, described by
\begin{eqnarray}
    ds^2&=&-f(r)dt^2+\frac{dr^2}{f(r)}+r^2(d\theta^2+sin^2\theta d\varphi^2)~,\label{metric1}\\
    f(r)&=&1-\frac{2M}{r}+\frac{Q^2}{r^2}+\frac{r^2}{L^2}~.
\end{eqnarray}
The Hawking temperature reads
\begin{align}
T=\frac{f'(r_h)}{4\pi}=\frac{1}{4\pi r_h}+\frac{3r_h}{4L^2\pi}-\frac{Q^2}{4\pi r_h^3},
\end{align}
where $r_h$ and $Q$ denote the radius of horizon and charge of the black holes. In the extended phase space, the AdS radius $L$ is related to the pressure as $P=3/(8\pi L^2)$~\cite{Kastor2009}, which leads to the following equation of state:
\begin{align}
P=\frac{T}{2r_h}-\frac{1}{8\pi r_h^2}+\frac{Q^2}{8\pi r_h^4}~.
\end{align}
By solving $(\partial_{r_h}P)_{T} = (\partial^2_{r_h}P)_{T} = 0$, the critical point ($P_c$, $T_c$, $r_{hc}$) = ($1/96\pi Q^2$, $\sqrt{6}/18\pi Q$, $\sqrt{6} Q$) is derived. Below this critical point, the isothermal curve demonstrates non-monotonic behavior, signifying a phase transition between small and large black holes. The coexistence curve is determined by applying the Maxwell equal area law to each isothermal curve \cite{Kubiznak2012}, as illustrated by the black solid curve in Fig.~\ref{pphaseDiagram}. However, above the critical point, distinguishing between small and large black holes becomes challenging, characterizing the supercritical region highlighted in the top right-hand corner in Fig.~\ref{pphaseDiagram}. Notably, the thermodynamic quantities: scaled variance $\Omega$, skewness $S\sigma$, and kurtosis $\kappa\sigma^2$ as defined in Eq.~(\ref{qqq}), can be employed to investigate the transition behavior between the small and large black hole-like subphases.

\begin{figure*}[!htbp]
\subfigure[]{\label{Omegasingle}\includegraphics[width=5.5cm]{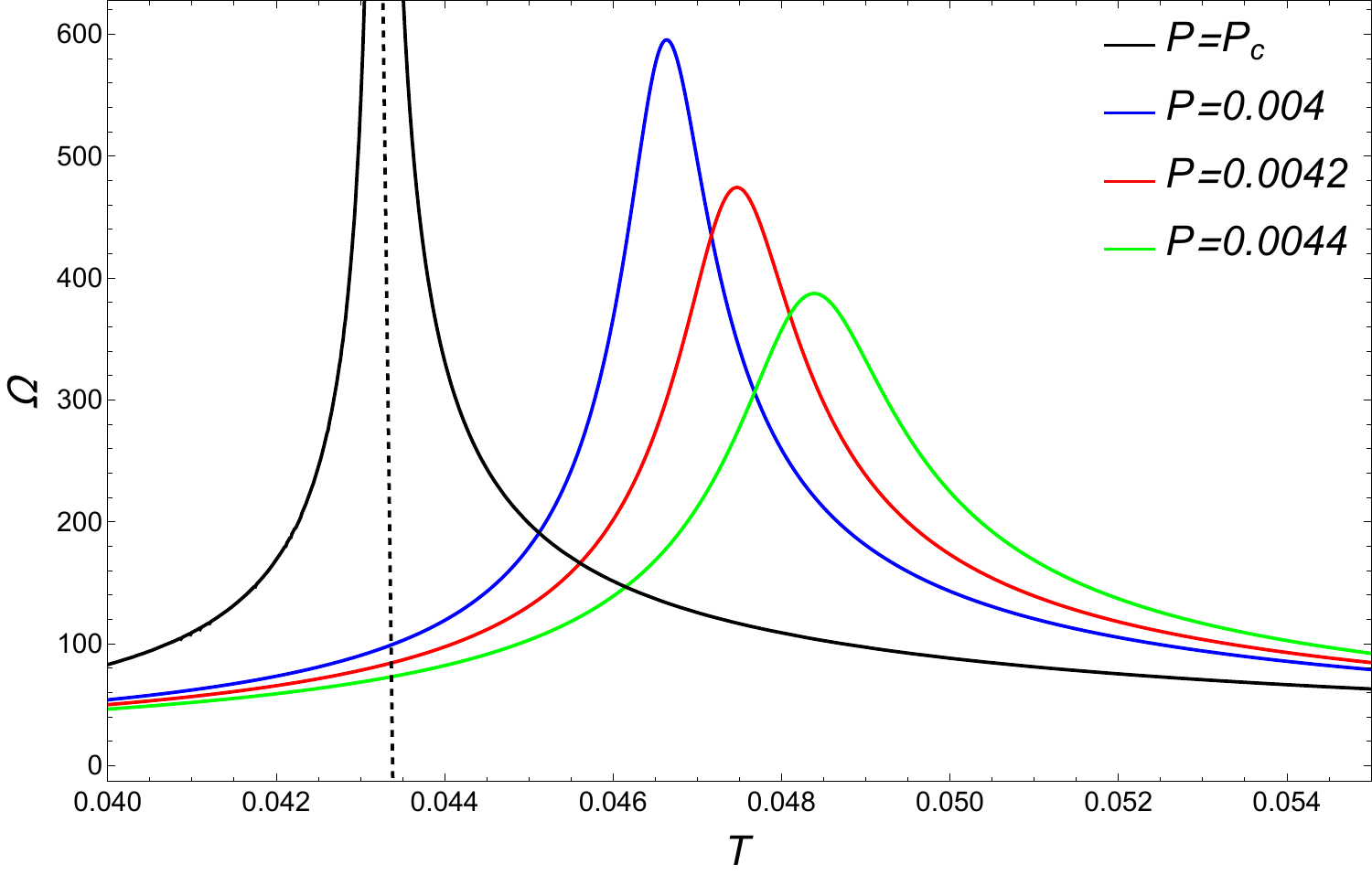}}
\subfigure[]{\label{Omega}\includegraphics[width=6.5cm]{Omega_1b.png}}
\subfigure[]{\label{Ssigma}\includegraphics[width=6.5cm]{Ssigma_1c.png}}
\subfigure[]{\label{KappaSigma}\includegraphics[width=6.5cm]{KappaSigma_1d.png}}
\caption{The behavior of the thermodynamic functions $\Omega$, $S\sigma$, and $\kappa\sigma^2$. (a) Behavior of the scaled variance $\Omega$ for different values of pressure. (b), (c), and (d) are the density plots of the scaled variance $\Omega$, skewness $S\sigma$, and kurtosis $\kappa\sigma^2$, respectively. The black hole charge $Q=1$. The black dashed line in (b) represents the Widom line. The cyan dashed lines indicate the points where $S\sigma$ and $\kappa\sigma^2$ are vanish in (c) and (d).}\label{pKappaden}
\end{figure*}

Following the gravitational path integral (\ref{action}), the Gibbs free energy can be obtained~\cite{Chamblin}
\begin{align}
 G=\frac{r_h}{4}+\frac{3Q^2}{4r_h}-\frac{2\pi P r_h^3}{3}~.
\end{align}
Further combining with the Hawking temperature, the scaled variance $\Omega$, skewness $S\sigma$, and kurtosis $\kappa\sigma^2$ can be obtained. Since they are in complicated forms, we only show their numerical results in Fig.~\ref{pKappaden}.

According to the definition of the Widom line, it represents the peak of the thermodynamic response function, which in this case corresponds to the peak of the scaled variance $\Omega$. To visualize this behavior clearly, we depict it in Fig.~\ref{Omegasingle} for $P \geq P_c$. Notably, when $P = P_c$, denoted in black, a divergent point emerges at the critical temperature. On either side of this point, $\Omega$ tends towards positive infinity. However, for $P > P_c$, the divergent behavior vanishes, showcasing only a peak. As pressure increases, this peak diminishes and shifts towards higher temperatures, aligning with the expected behavior of the Widom line. The presence of this peak divides each curve into two branches: one with a positive slope representing the small black hole-like phases, and the other with a negative slope corresponding to the large black hole-like phases, akin to the behavior observed in VdW fluids.

Furthermore, we present the density plots of $\Omega$ in Fig.~\ref{Omega} within the $P$-$T$ parameter space. The figure illustrates that, in most regions of the parameter space, $\Omega$ remains close to zero. However, beyond the critical point, $\Omega$ exhibits significantly large values before decreasing as the distance from the critical point increases. The dashed black line, representing the Widom line, originates from the critical points and extends towards high temperatures and pressures, effectively dividing the supercritical black hole phase into small and large black hole-like regions. If we hypothesize that the value of $\Omega$ can serve as an indicator for distinguishing between small and large black hole-like phases, differentiation between these phases near the critical point becomes straightforward.

The density plots for skewness $S\sigma$ and kurtosis $\kappa\sigma^2$ are also depicted in Figs.~\ref{Ssigma} and \ref{KappaSigma}. Unlike $\Omega$, their values can vary between positive and negative, serving as additional tools to differentiate between small and large black hole-like phases. It is proposed that the sign of $S\sigma$ could be utilized to discern these subphases within the supercritical region. The zero value line marked in cyan dashed is illustrated in Fig.~\ref{Ssigma}. Upon comparison with the lines presented in Fig.~\ref{Omega}, it becomes apparent that near the critical point, they align, yet noticeable deviations emerge at higher temperatures and pressures. In Fig. \ref{KappaSigma}, two zero value lines for $\kappa\sigma^2$ are observed. The negative value region is enclosed by these lines, while in other parameter regions, $\kappa\sigma^2$ assumes positive values. Previous studies such as Ref.~\cite{Stephanov2011} suggest that these lines correspond to symmetry lines in the QCD system. 
In addition, we note that in recent studies~\cite{Wang:2025ctk,Jin:2025sub}, the authors compared the critical behavior of supercritical black holes and supercritical water, which exhibits the same critical exponent of $1.5$. Inspired by their findings, we also calculated the critical exponent of $\kappa\sigma^2$. Based on our results, the fitted critical behavior near the critical point is as follows
\begin{align}
    \delta p\approx0.0068(\frac{T}{T_c}-1)^{1.4942}~,
\end{align}
in which, $\delta p$ denotes the difference between the two symmetry lines in Figs.~\ref{KappaSigma}. Our results show close agreement with the critical exponent of $1.5$ near the critical point.

Through the aforementioned study, we ascertain that the Widom line, associated with the maximal $\Omega$, serves as an effective means to differentiate between the small and large black hole-like subphases within the supercritical region.

{\emph{3. Frenkel line and dynamic crossover}.} Shifting our focus to the Frenkel line, it offers a distinct criterion for discerning between the small and large black hole-like subphases from a dynamic perspective.

Generally, when a classical system undergoes a perturbation, it gradually returns to equilibrium over time, and the duration needed for this restoration is known as the relaxation time. Various physical states exhibit distinct relaxation times. For instance, liquids require more time than gases to reach equilibrium, while solids necessitate even longer durations. This disparity arises from the higher molecular density and shear rigidity in liquids and solids, leading to prolonged relaxation times.

Alternatively, another function that can highlight these distinctions is the velocity autocorrelation function (VAF)~\cite{Brazhkin2013}. The VAF for gases exhibits a monotonic  decay, whereas for liquids and solids, it decays in an oscillatory fashion. The transitions between monotonic and oscillatory decay patterns align with the Frenkel line. Analogous to the Widom line, the Frenkel line can persist in the supercritical region, providing an avenue to scrutinize the thermodynamic characteristics of black holes.

For our purpose, we devote our attention to the QNMs of the black hole, which quantify the decay rate of perturbations. Specifically, our attention is directed towards scalar perturbations, which are dictated by the massless Klein-Gordon equation,
\begin{align}
    \frac{1}{\sqrt{-g}}\partial_\mu(\sqrt{-g}g^{\mu\nu}\partial_\nu\Phi)=0~.
\end{align}\label{metric2}
Adopting the following decomposition:
\begin{align}
    \Phi=\frac{1}{r}\psi(r)Y(\theta,\varphi)e^{-i\omega v}~,
\end{align}\label{metric2}
one shall obtain the perturbation equation
\begin{align}
f(r)\frac{d^2\psi(r)}{dr^2}+(f'(r)-2i\omega)\frac{d\psi(r)}{dr}-V(r)\psi(r)=0~,
\end{align}
with
\begin{align}
V(r)=\frac{f'(r)}{r}+\frac{l(l+1)}{r^2}~.
\end{align}
Without loss of generality, we shall set the angular quantum number $l$ to zero. In principle, we should consider all channels of QNMs of black holes to fully characterize their dynamics. However, as a preliminary investigation, we focus on the simplest channel of a probing massless scalar. By imposing the ingoing boundary condition at the event horizon and requiring $\Phi$ to vanish at the asymptotic infinity boundary, the frequencies of the QNMs can be determined using the Horowitz-Hubeny method~\cite{Horowitz2000}. For our study, we employ the Chebyshev spectral method in conjunction with the Newton-Raphson relaxation technique~\cite{Horowitz:2011dz} to numerically compute these frequencies.

\begin{figure}[t]
\center
\includegraphics[width=1\columnwidth]{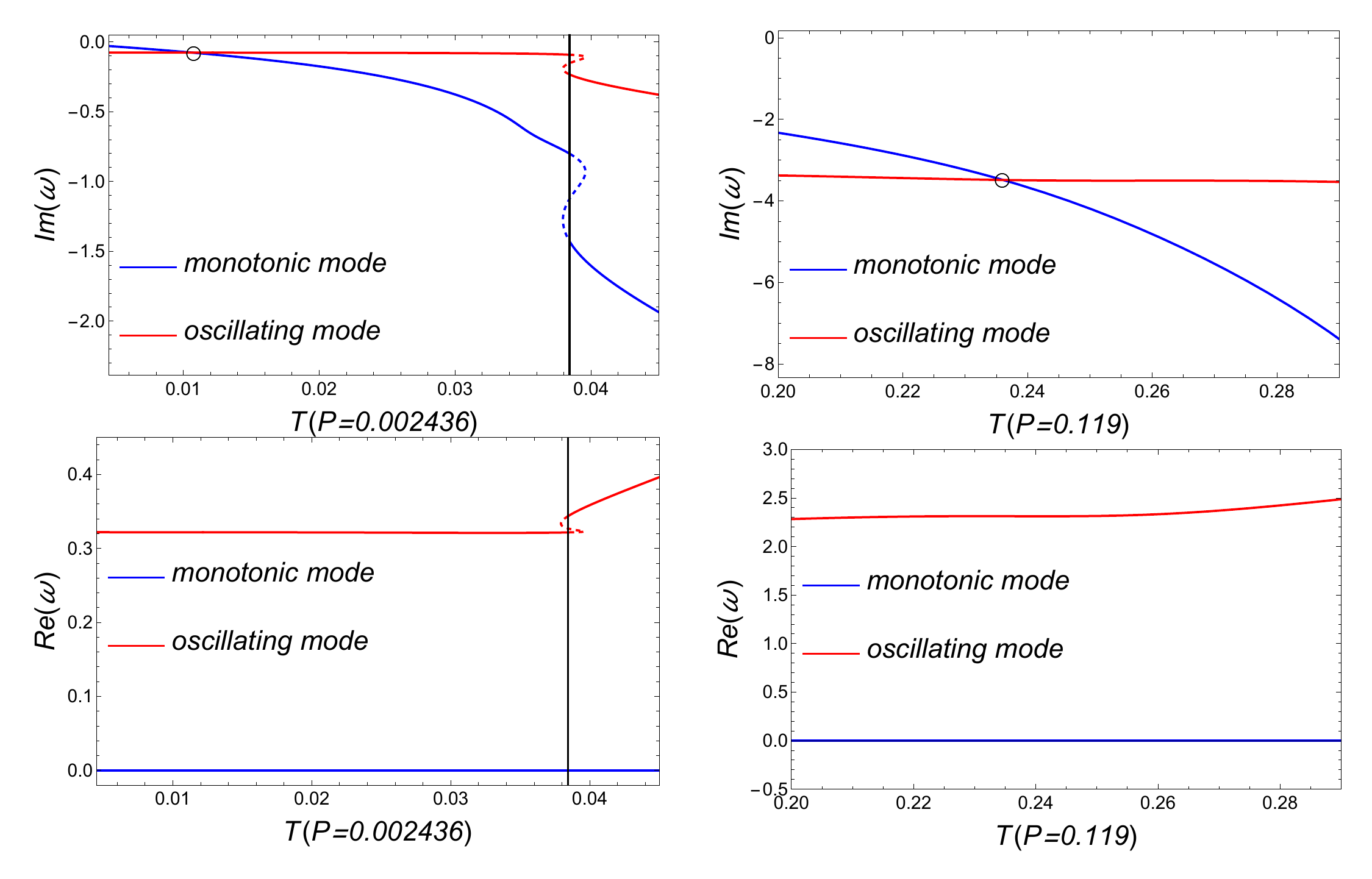}
\caption{The QNMs as a function of temperature at fixed pressures. The blue and red lines correspond to the monotonic and oscillatory modes, respectively. Here, the solid lines represent thermodynamically stable states, while the dashed lines represent thermodynamically metastable or unstable states. The black lines represent the phase transition point of first-order phase transition.
}\label{pQNM2}
\end{figure}

In Fig.~\ref{pQNM2}, we depict the relationship between QNMs and temperature for both scenarios.  The left panel corresponds to the first-order phase transition region, while the right panel represents the supercritical region. As the black hole temperature increases, the blue-marked monotonic mode swiftly shifts downward, while the red-marked oscillatory mode displays minimal alteration. At a specific temperature, the lowest mode transitions from the monotonic to the oscillatory type. This transition point, marked by a black circle, signifies a dynamic crossover in the system.
The Frenkel line, represented by a blue dashed curve, has been calculated and is shown in Fig.~\ref{pphaseDiagram}.

From Fig.~\ref{pQNM2}, it is observed that the real and imaginary components of the QNMs exhibit non-monotonic behaviors below the critical pressure, while displaying solely monotonic behaviors above the critical pressure. This observation can be used to probe the first-order phase transition to a certain extent. Furthermore, irrespective of temperature and pressure, the crossover behavior persists. Notably, the Frenkel line does not intersect the critical point, which is in contrast to the Widom line. This result implies that the behaviors between the small and large black hole-like subphases differ concerning both thermodynamic and dynamic aspects.

{\emph{4. Conclusions}}. This study delves into the crossover phenomena within the thermodynamic supercritical region of charged AdS black holes, focusing on the distinctive Widom line and Frenkel line delineated in the phase diagram (Fig.~\ref{pphaseDiagram}) by the black and blue dashed lines, respectively. The results illustrate the segregation of small and large black hole-like subphases by these lines, providing a potential avenue to gain deeper insights into the thermodynamic nature of black holes that extend beyond traditional phase transitions.

Dynamically, we examined the QNMs spectrum concerning the scalar perturbations and identified the Frenkel line, a delineation demarcating the transition between monotonic and oscillatory decay patterns in perturbative relaxation. This dynamic crossover, unlike the Widom line, remains outside the critical point, yet endures in the supercritical zone, unveiling a distinctive shift in the relaxation dynamics of black holes.


The Widom line and the Frenkel line originate from two distinct criteria, thermodynamics and dynamics, respectively. Although both can differentiate subphases in the supercritical region, they exhibit distinct characteristics. The key difference between these two demarcation lines is that the Frenkel line does not stem from the critical point. In contrast, the Widom line is derived from higher-order derivatives of the free energy and thus always originates from the critical point.

We note that some recent studies have also investigated crossover behaviors in supercritical black holes. The authors of Ref.~\cite{Xu:2025jrk} obtained the Widom line using the Lee-Yang phase transition theory, with results showing good agreement with ours. Reference \cite{Li:2025tdd} focuses on dynamical crossovers, while Ref.~\cite{Wang:2025ctk} emphasizes universal scaling laws. We derive the supercritical thermodynamic and dynamic crossovers from the most commonly used physical quantities describing black hole physics (black hole thermodynamics and QNMs), which allows us to obtain supercritical information more conveniently.

Our study establishes a connection between black hole thermodynamics and universal crossovers observed in both classical and quantum systems within the supercritical domain. The recognition of the Widom and Frenkel lines within charged AdS black holes underscores a profound analogy between gravitational systems and conventional matter, presenting a unified framework for exploring supercritical phenomena. Subsequent studies could expand upon this foundation by exploring rotating black holes and other black hole solutions.

Moreover, exploring the complete time-dependent evolution of black holes undergoing first-order phase transitions in the extended phase space, while maintaining dynamical and thermodynamic stability, presents intriguing issues. This approach aligns with the results observed in holographic superfluid systems via the AdS/CFT correspondence. In particular, investigation between the black hole thermodynamics and dynamic relaxation has the potential to unveil profound correlations between criticality and black hole stability, with implications extending to gravitational wave astronomy and quantum gravity research.

{\emph{Acknowledgements}.} This work was supported by the National Natural Science Foundation of China (Grant Nos. 12473001, 11975072, 11875102, 11835009, and 11965013), the National SKA Program of China (Grant Nos. 2022SKA0110200 and 2022SKA0110203), and the National 111 Project (Grant No. B16009). ZYN is partially supported by Yunnan High-level Talent Training Support Plan Young $\&$ Elite Talents Project (Grant No. YNWR-QNBJ-2018-181).

\end{CJK*}
\bibliographystyle{apsrev4-1}
\bibliography{reference}

\end{document}